\documentclass[epj]{webofc}
\usepackage[varg]{txfonts}   
\wocname{EPJ Web of Conferences}
\woctitle{CONF12}

\usepackage{amsmath,amssymb}
\usepackage{dsfont}
\usepackage{slashed}
\usepackage{verbatim}    
\usepackage{graphicx} 

\newcommand{\be}{\begin{equation}}
\newcommand{\ee}{\end{equation}}
\newcommand{\bea}{\begin{eqnarray}} 
\newcommand{\eea}{\end{eqnarray}}

\newcommand{\MSbar}{{\overline{\rm MS}}}
\newcommand{\pa}{\partial}

\newcommand{\qslash}{{\not{\hspace{-0.08cm}q}}}
\newcommand{\thb}{\bar{\theta}}
\newcommand{\la}{\lambda}
\newcommand{\si}{\sigma}
\newcommand{\al}{\alpha}

\def\openone{\leavevmode\hbox{\small1\kern-4pt\normalsize1}}

\begin{document}
\selectlanguage{english}
\title{One-loop calculations in Supersymmetric Lattice QCD}

\author{M.~Costa\inst{1}\fnsep\thanks{\email{kosta.marios@ucy.ac.cy}} \and
        H.~Panagopoulos\inst{1}
}

\institute{Department of Physics, University of Cyprus, CY-1678 Nicosia, Cyprus}

\abstract{ 
We study the self energies of all particles which appear in a lattice regularization of supersymmetric QCD (${\cal N}=1$). We compute, perturbatively to one-loop, the relevant two-point Green's functions using both the dimensional and the lattice regularizations. Our lattice formulation employs the Wilson fermion acrion for the gluino and quark fields. The gauge group that we consider is $SU(N_c)$ while the number of colors, $N_c$ and the number of flavors, $N_f$, are kept as generic parameters. We have also searched for relations among the propagators which are computed from our one-loop results. We have obtained analytic expressions for the renormalization functions of the quark field ($Z_\psi$), gluon field ($Z_u$), gluino field ($Z_\lambda$) and squark field ($Z_{A_\pm}$). 

We present here results from dimensional regularization, relegating to a forthcoming publication \cite{HP:2016} our results along with a more complete list of references. Part of the lattice study regards also the renormalization of quark bilinear operators which, unlike the non-supersymmetric case, exhibit a rich pattern of operator mixing at the quantum level. 
}
\maketitle
\section{Introduction}
In recent years, methods of extracting nonperturbative information for Supersymmetric Theories through lattice simulations are being studied extensively \cite{Feo,Catterall:2011,Kaplan:2009}, from a number of viewpoints. There are a number of important physical questions regarding SUSY to be ultimately addressed on the lattice, such as the nature of SUSY breaking, and the phase diagram of SUSY models. Such questions have become increasingly relevant in recent years, in the context of investigations of BSM Physics. Many notorious problems arise when formulating SUSY models on the lattice \cite{Catterall:2014}, such as the emergence of a plethora of counterterms in the action and the need for fine-tuning of masses and coupling constants \cite{Jack&Jones,Taniguchi:2000,Giet&Poppitz}. The present work investigates these problems using, as a representative nontrivial model, supersymmetric ${\cal N}=1$ Quantum Chromodynamics (SQCD).

The fundamental fields involved in the construction of the SQCD Lagrangian are well-known and we present them below for completeness. There are two types of fields, chiral superfields and vector superfields \cite{Weinberg:1980bf,Wess&Bagger:1992}. The physical components of a chiral superfield $\Phi$, are the matter fields: $A$ which represents a scalar boson (squark), $\psi$ which is a two-component spinor (quark - spin  $\frac{1}{2}$) and $F$ which is an auxiliary complex scalar field. In superspace notation ($x$: spacetime coordinates, $\theta$/$\thb$: anticommuting coordinates) the chiral superfield $\Phi$ in terms of the above component fields is:
\bea
    \Phi(x ,\theta ,\thb) &=& A(x) + \sqrt{2} \, \theta \psi(x)  +  \theta \theta \, F(x) +
 i \theta \, \sigma^{\mu} \, \thb  \, \pa_{\mu} A(x) \\ \nonumber
& + & \frac{i}{\sqrt{2}} \theta \theta\, \thb \, \bar\sigma^{\mu} \, \pa_{\mu} \psi(x) + \frac{1}{4} \; \theta \theta \, \thb \thb \, \pa_{\mu} \pa^{\mu} A(x)
\eea 
The general form of a vector superfield $V(x,\theta, \thb)$ is:
\bea
V(x,\theta,\thb) &=&  C(x) +  i\theta \chi(x) -  i \thb \bar{\chi}(x)  +  \frac{i}{2} \theta \theta \, \big[M(x) \, + \, iN(x) \big] -  \frac{i}{2} \thb \thb \big[M(x) \, - \, iN(x) \big] \\ \nonumber
& - & \theta \, \sigma^{\mu} \, \thb \, u_{\mu}(x)  +  i \theta \theta \, \thb \left[ \bar{\la}(x) \, + \, \frac{i}{2} \bar{\si}^{\mu} \pa_{\mu} \chi(x) \right] -  i\thb \thb \, \theta \left[\la(x) \, + \, \frac{i}{2} \sigma^{\mu} \pa_{\mu} \bar{\chi}(x) \right] \\ \nonumber 
& + & \frac{1}{2} \, \theta \theta \, \thb \thb \, \left[D(x) \, + \, \frac{1}{2} \pa_{\mu} \pa^{\mu} C(x) \right] \ .
\eea
We can choose a special gauge where the components $C , \chi , M ,N$ are zero. This defines the Wess$-$Zumino (WZ) gauge. A vector superfield in Wess$-$Zumino gauge reduces to the form:
\be    
V(x,\theta, \thb) = -\theta \, \si^{\mu} \, \thb \, u_{\mu}(x)  + i \theta \theta \, \thb \bar{\la}(x) - i \thb \thb \,\theta \la(x) + \ \frac{1}{2} \, \theta \theta \, \thb \thb \, D(x) \ . 
\ee
where $u_{\mu} = u_{\mu}^{(\al)} T^{(\al)}$ is the gluon field, $\la = \la^{(\al)} T^{(\al)}$ is the gluino field and $D= D^{(\al)} T^{(\al)}$ is an auxiliary real scalar field; $(\al)$ is a color index in the adjoint representation of the gauge group, and the generators $T^{(\al)}$ satisfy ${\rm Tr} (T^{\alpha}T^{\beta}) = 1/2 \,\delta^{\alpha\,\beta}$.

The minimal content of the SQCD Lagrangian, in order to contain gluons and 4-component massive quarks is two chiral super$-$multiplets $\Phi_+ = \{A_+,\psi_+, F_+\}$ and $\Phi_- = \{A_-,\psi_-, F_-\}$ and a vector super$-$multiplet $V = \{u_{\mu}, \la, D\}$. We combine the components $\psi_+$, $\psi_-$ to create Dirac spinors. In order to obtain a renormalizable theory, we need to construct a Lagrangian with products of superfields having dimensionality $\leq 4$; in addition, we require Lorentz invariance as well as invariance under supersymmetric gauge transformations:
\be
\Phi'_+ = e^{-i \Lambda} \Phi_+\,, \,\,\,\,\Phi'_- =  \Phi_- e^{i \Lambda},\,\,\,\,e^{V'} = e^{-i \Lambda^{\dagger}} e^{V} e^{i \Lambda},
\ee
where $\Lambda$ is an arbitrary chiral superfield. The Lagrangian takes the form:
\be
{\cal L} = \frac{1}{8\,g} {\rm Tr} \big( W^{\al} \, W_{\al}|_{\theta \theta} +\bar{W}_{\dot{\al}} \, \bar{W}^{\dot{\al}}|_{\thb \thb}\big) 
+\big( \Phi^{\dagger}_+ \, e^V \, \Phi_+ +  \Phi_- \, e^{-V} \, \Phi^{\dagger}_- \big)|_{\theta \theta \thb \thb}
+ m \big(\Phi_-\Phi_+|_{\theta \theta} +  \Phi^{\dagger}_+\Phi^{\dagger}_-|_{\thb \thb} \big)
\label{superfieldLag}
\ee
where $W_{\al} =  -\frac{1}{4} \bar{{\cal D}}\bar{{\cal D}} \,e^{-V} \,  {\cal D}_{\al} \,e^{V} $ is the supersymmetric field strength and  the supersymmetric covariant derivative is defined as: ${\cal D}_{\al} =  \frac{\pa}{\pa \theta^{\al}}\ + \ i \si^{\mu}_{\al \dot{\al}} \, \thb^{\dot{\al}} \, \pa_{\mu} \,\,, \bar{{\cal D}}_{\dot{\al}} =  - \frac{\pa}{\pa \thb^{\dot{\al}}} - \ i \theta^{\al} \, (\si^{\mu})_{\al \dot{\al}} \, \pa_{\mu}$. Combining the components of $\Phi_+$ with $\Phi_-$ we can construct a 4 component Dirac Spinor ($\psi_D$).
Upon functionally integrating over the auxiliary fields, rescaling $V \rightarrow   2\,g\,V$ and after a Wick rotation, the form of the Euclidean action in 4 dimensions in Dirac notation ${\cal S}^E_{\rm SQCD}$ is:\\
\bea
{\cal S}^E_{\rm SQCD} & = & \int d^4x \Big[ \frac{1}{4}u_{\mu \nu}^{(\alpha)} u_{\mu \nu}^{(\alpha)} + \frac{1}{2} \bar \lambda^{(\alpha)}_M \gamma^E_\mu {\cal{D}}_\mu\lambda^{(\alpha)}_M \nonumber \\
&+& {\cal{D}}_\mu A_+^{\dagger}{\cal{D}}_\mu A_+ + {\cal{D}}_\mu A_- {\cal{D}}_\mu A_-^{\dagger}+ \bar \psi_D \gamma^E_\mu {\cal{D}}_\mu \psi_D \nonumber \\
&+&i \sqrt2 g \big( A^{\dagger}_+ \bar{\lambda}^{(\alpha)}_M T^{(\alpha)} P_+^E \psi_D  -  \bar{\psi}_D P_-^E \lambda^{(\alpha)}_M  T^{(\alpha)} A_+ +  A_- \bar{\lambda}^{(\alpha)}_M T^{(\alpha)} P_-^E \psi_D  -  \bar{\psi}_D P_+^E \lambda^{(\alpha)}_M  T^{(\alpha)} A_-^{\dagger}\big)\nonumber\\  
&+& \frac{1}{2} g^2 (A^{\dagger}_+ T^{(\alpha)} A_+ -  A_- T^{(\alpha)} A^{\dagger}_-)^2 - m ( \bar \psi_D \psi_D - m A^{\dagger}_+ A_+  - m A_- A^{\dagger}_-)\Big] \,. 
\label{susylagr}
\eea
where $ \la_M= \left( {\begin{array}{c} \la_a\\ \bar \la^{\dot a} \end{array} } \right)$ and
   $\psi_D^T = \left( {\begin{array}{c} \psi_{+ a}\\ \bar\psi_-^{\dot a} \end{array}} \right)$,\, $P_\pm^E= \frac{1 \pm \,\gamma_5^E}{2}$ , $\gamma_5^E = \gamma_0^E \gamma_1^E \gamma_2^E \gamma_3^E$. ${\cal S}^E_{\rm SQCD}$ is invariant under supersymmetric transformations ($\xi_M$: Majorana spinor parameter):
\bea
\delta_\xi A_+ & = & - \sqrt2  \bar\xi_M P_+^E \psi_D \, , \nonumber \\
\delta_\xi A_- & = & - \sqrt2  \bar\psi_D P_+^E \xi_M  \, , \nonumber \\
\delta_\xi (P_+^E \psi_{D}) & = & \sqrt2 ({\cal{D}}_\mu A_+) P_+^E \gamma_\mu^E \xi_M  - \sqrt2 m P_+^E \xi_M A_-^{\dagger}\, , \nonumber \\
\delta_\xi (P_-^E \psi_D) & = &   \sqrt2 ({\cal{D}}_\mu A_-)^{\dagger} P_-^E \gamma_\mu^E \xi_M  - \sqrt2 m  A_+ P_-^E \xi_M\, ,\nonumber \\
\delta_\xi u_\mu^{(\alpha)} & = & - \bar \xi_M \gamma_\mu^E \lambda^{(\alpha)}_M, \nonumber \\
\delta_\xi \lambda^{(\alpha)}_M & = & \frac{1}{4} u_{\mu \nu}^{(\alpha)} [\gamma_{\mu}^E,\gamma_{\nu}^E] \xi_M - 2 i g \gamma_5^E \xi_M (A^{\dagger}_+ T^{(\alpha)} A_+ -  A_- T^{(\alpha)} A^{\dagger}_-)\,.
\label{susytransfDiracE}
\eea
\section{The Calculation}
\label{calculation}
We calculate perturbatively a number of 2-pt Green's functions up to one-loop, both in the continuum and on the lattice. The quantities that we study are the self energies of the quark ($\psi$), gluon ($u_\mu$), squark ($A$) and gluino ($\lambda$) fields, as well as the 2-pt Green's functions of the quantities ${\cal O}_i(x)=\bar \psi(x) \Gamma_i \psi(x)$, using both dimensional regularization (DR) and lattice regularization (L). The index "$i$'' refers to the possibilities of the gamma matrices: (scalar) $\Gamma_S= \openone$, (pseudoscalar) $\Gamma_P=\gamma_5$, (vector) $\Gamma_V =\gamma_\mu$, (axial) $\Gamma_A=\gamma_5\gamma_\mu$, (tensor) $\Gamma_T= \gamma_{\mu}\gamma_{\nu}$.

The first step in our perturbative procedure is to calculate the 2-pt Green's functions in the continuum where we regularize the theory in D-dimensions ($D=4-2\,\epsilon$). The continuum calculations are necessary in order to compute the $\MSbar$-renormalized Green's functions; these enter the subsequent calculation of the corresponding Green's functions using lattice regularization and $\MSbar$ renormalization. The continuum results also provide the renormalization functions of the quark field ($Z_\psi$), squark field ($Z_{A_\pm}$), gluon field ($Z_u$) and gluino field ($Z_\lambda$) and of the complete set of quark bilinear operators ${\cal{O}}_i$ ($Z_i$). For the extraction of the renormalization functions, we applied the $\MSbar$ scheme at a scale $\bar{\mu}$. Once we have computed the renormalization functions in the $\MSbar$ scheme we can construct their RI$'$ counterparts using conversion factors which are immediately extracted from our computation to the required perturbative order. Being regularization independent, these same conversion factors can then be also used for lattice renormalization functions.

\begin{figure}[ht!]
\centering
\includegraphics[scale=0.225]{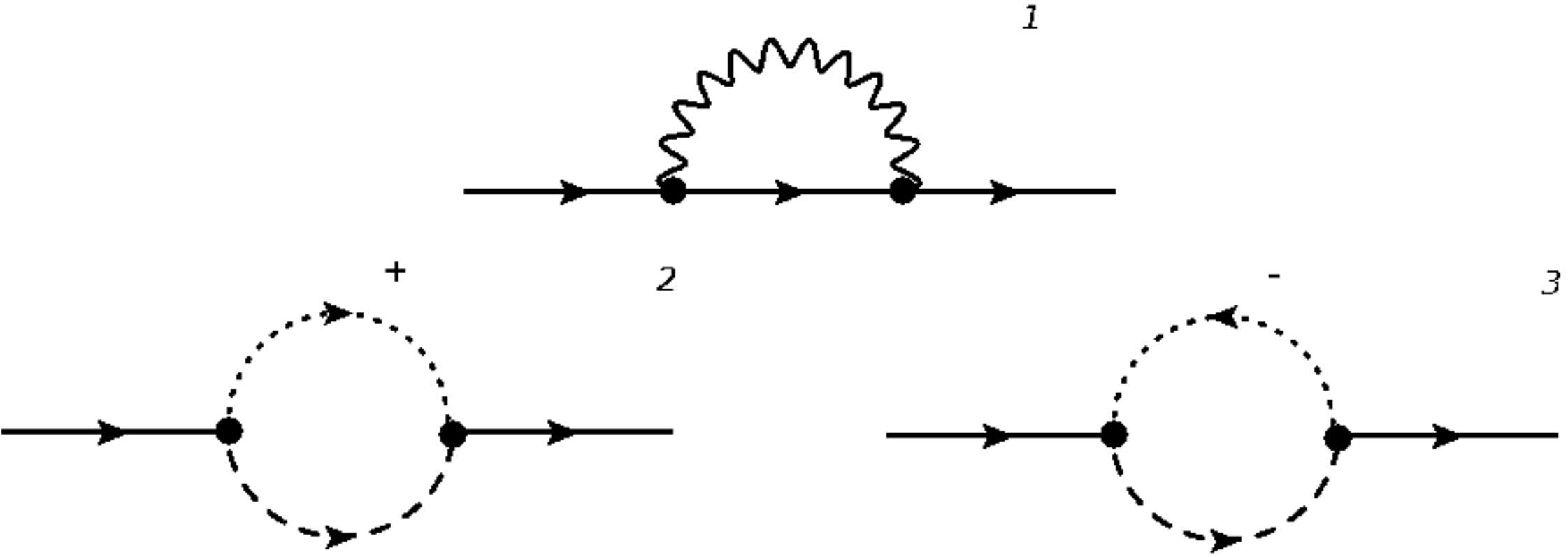}
\caption{One-loop Feynman diagrams contributing to the 2-pt Green's function of the quark propagator, $\langle \psi(x) \bar \psi(y) \rangle$.}
\label{quark2pt}
\end{figure}

\begin{figure}[ht!]
\centering
\includegraphics[scale=0.225]{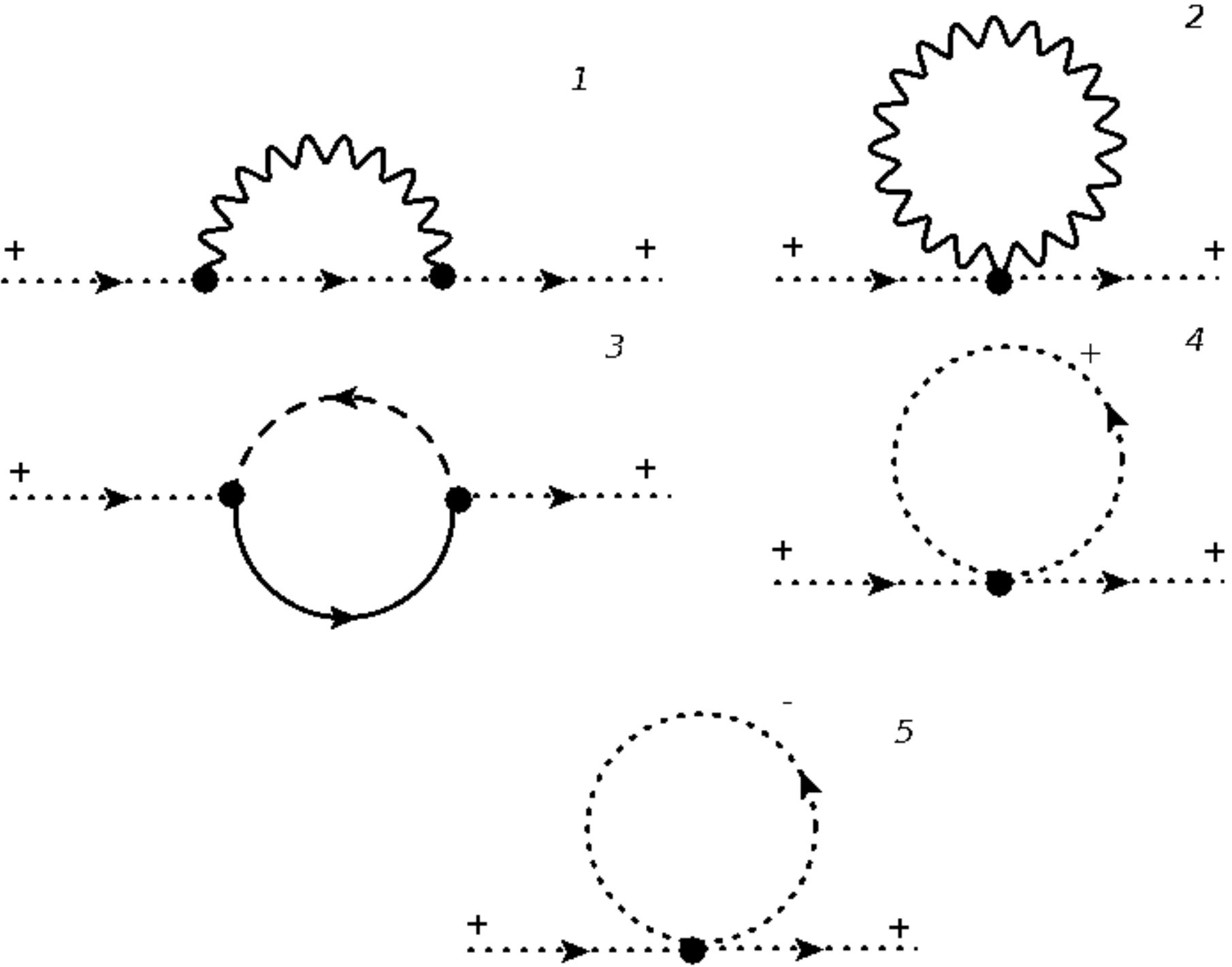}
\caption{One-loop Feynman diagrams contributing to the 2-pt Green's function of the squark propagator, $\langle A_+(x) A_+^{\dagger}(y) \rangle$. Similar diagrams apply to the propagator of the $A_-$ field.
  }
\label{squark2pt}
\end{figure}

\begin{figure}[ht!]
\centering
\includegraphics[scale=0.225]{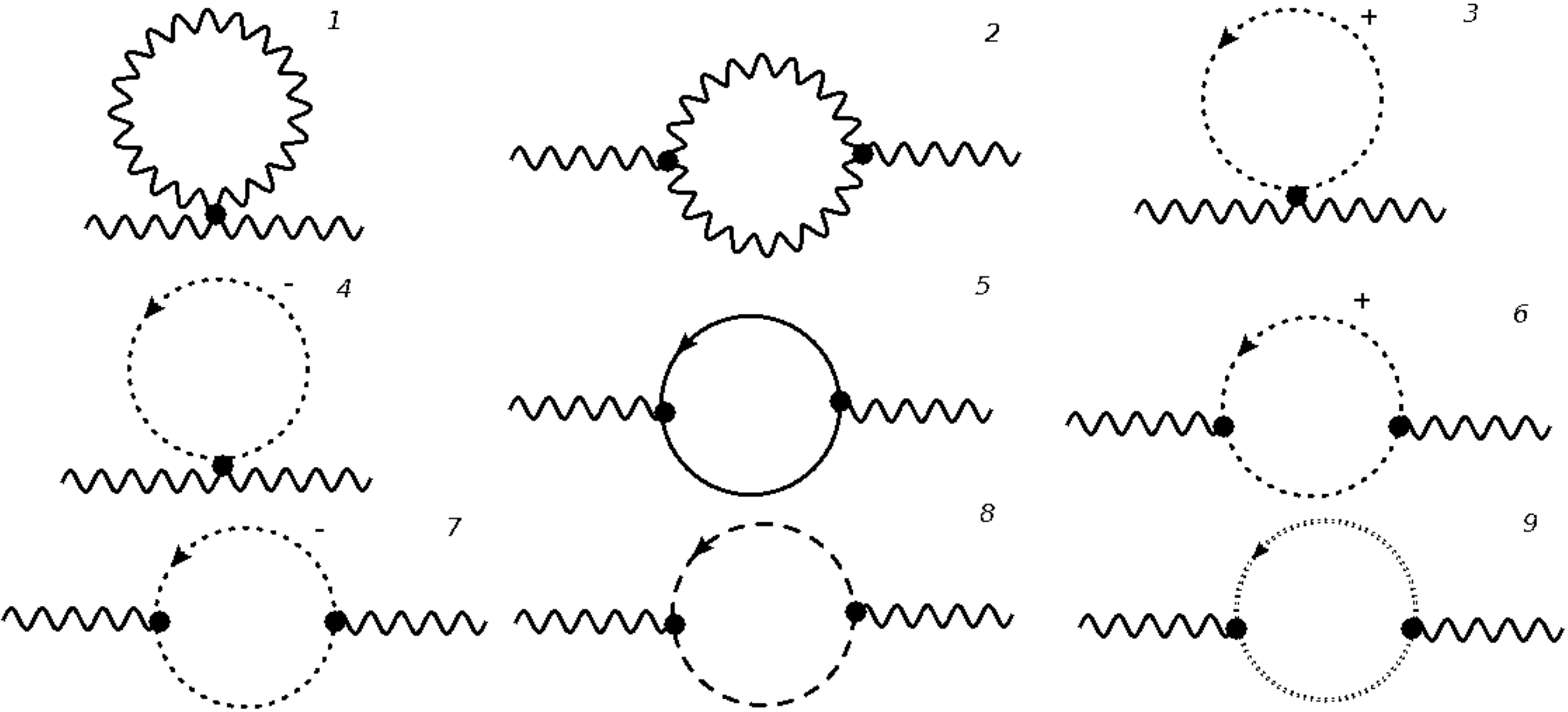}
\caption{One-loop Feynman diagrams contributing to the 2-pt Green's function of the gluon propagator, $\langle  u_\mu^{(\alpha)}(x) u_\nu^{(\beta)}(y) \rangle$. The last Feynman diagram is the one with a closed ghost loop coming from the ghost part of the action.
  }
\label{gluon2pt}
\end{figure}

\begin{figure}[ht!]
\centering
\includegraphics[scale=0.225]{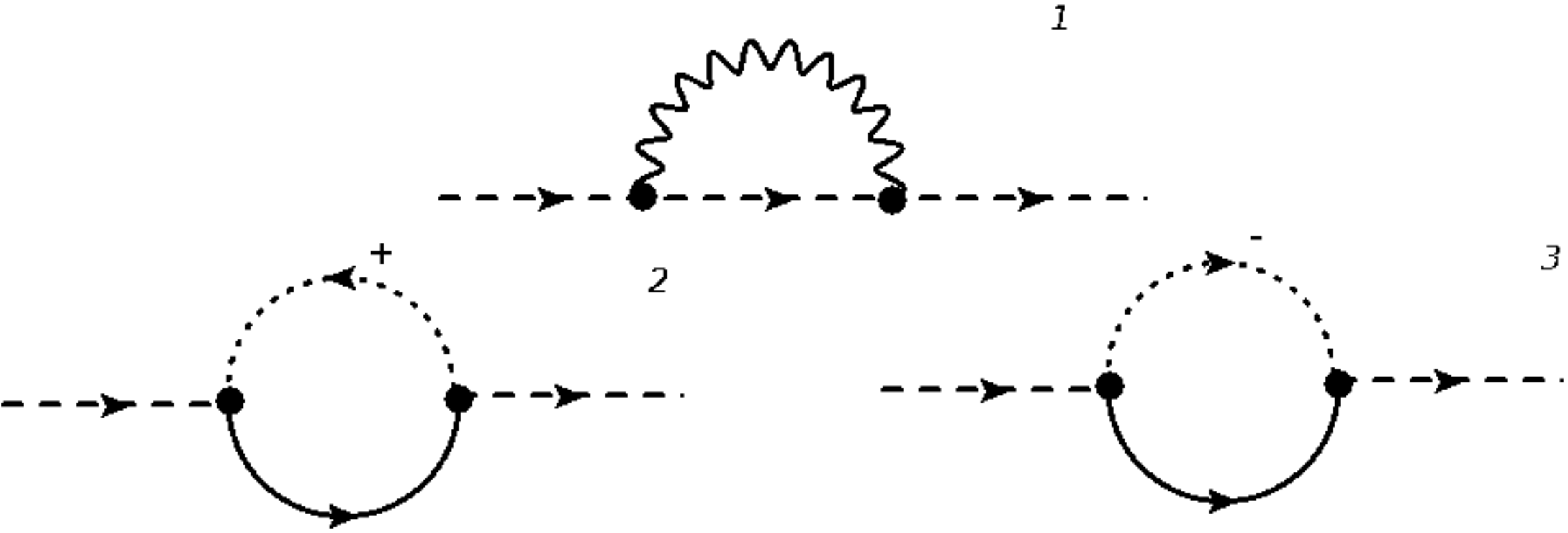}
\caption{One-loop Feynman diagrams contributing to the 2-pt Green's function of the gluino propagator, 
$\langle  \lambda^\alpha(x) \bar \lambda^\beta(y) \rangle$.
  }
\label{gluino2pt}
\end{figure}

\begin{figure}[ht!]
\centering
\includegraphics[scale=0.225]{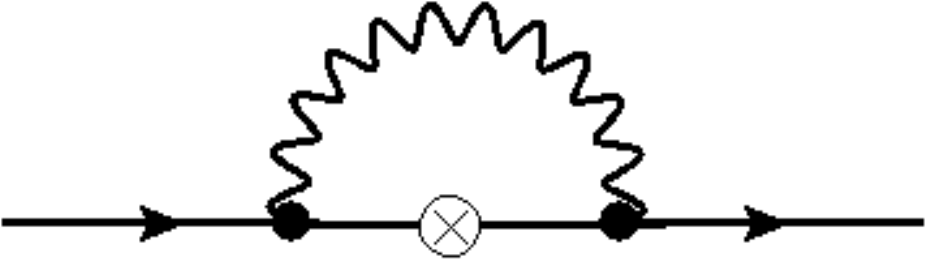}
\caption{One-loop Feynman diagram contributing to the 2-pt Green's function of $\langle \psi(x) {\cal O}_i(z) \bar \psi(y) \rangle$, where ${\cal O}_i$ are all local quark bilinear operators. 
The circled cross denotes the quark bilinear operator insertion.
}
\label{quarkbilinear}
\end{figure}

The aforementioned renormalization functions are defined as follows:
\bea
\psi^R=\sqrt{Z_\psi}\,\psi^B,\,\,\, A^R_\pm=\sqrt{Z_{A_\pm}}\,A^B_\pm,\,\,\, u_{\mu}^R=\sqrt{Z_u}\,u^B_{\mu},\,\,\, \la^R=\sqrt{Z_\la}\,\la^B,\,\,\, {\cal{O}}_i^R=Z_i\,{\cal{O}}_i^B.
\eea
The one-loop Feynman diagrams (one-particle irreducible (1PI)) contributing to $\langle \psi(x) \bar \psi(y) \rangle$  are shown in Fig.~\ref{quark2pt}, those contributing to $\langle A(x) A^{\dagger}(y) \rangle$ in Fig.~\ref{squark2pt}. One-loop Feynman diagrams contributing to the Green's function $\langle  u_\mu^{(\alpha)}(x) u_\nu^{(\beta)}(y) \rangle$ and $\langle  \lambda^{(\alpha)}(x) \bar \lambda^{(\beta)}(y) \rangle$ are shown in Fig.~\ref{gluon2pt} and Fig.~\ref{gluino2pt}, respectively. The Feynman diagram that enters the calculation of the Green's function $\langle  \psi(x) {\cal O}_i(z) \bar \psi(y) \rangle $ up to one-loop is shown in Fig.~\ref{quarkbilinear}. 

\section{Continuum Results}
\label{calculation}
Here we can collect all our results for the 2-pt Green's functions \footnote{A Fourier transformation, with momentum $q$, is intended for all Green's functions appearing on the lhs of Eqs.~(\ref{GF2quark}-\ref{GF2tensor}) and~(\ref{GF2quarklatt}).}:
\be
\langle \psi(x) \bar \psi(y) \rangle^{DR}_{\rm{amp}}\Big{|}_{m=0}= i \qslash \left[ 1 - \frac{g^2\,C_F}{16\,\pi^2} \left( \frac{2+\alpha}{\epsilon} + 4 + \alpha +  (2+\alpha) \log\left(\frac{\bar\mu^2}{q^2} \right) \right) \right]
\label{GF2quark}
\ee
where $C_F=(N_c^2-1)/(2\,N_c)$ is the quadratic Casimir operator in the fundamental representation, $\alpha$ is the gauge parameter ($\alpha=1(0)$ corresponds to the Feynman (Landau) gauge). A Kronecker delta for color indices is understood in Eqs.~(\ref{GF2quark}) and~(\ref{GF2squark}).
\be
\langle A_\pm(x) A_\pm^{\dagger}(y) \rangle^{DR}_{\rm{amp}}\Big{|}_{m=0}= q^2 \left[ 1 - \frac{g^2\,C_F}{16\,\pi^2} \left( \frac{1+\alpha}{\epsilon} + \frac{16}{3} +  (1+\alpha) \log\left(\frac{\bar\mu^2}{q^2} \right) \right) \right]
\label{GF2squark}
\ee
\bea
\langle  u_\mu^{(\alpha)}(x) u_\nu^{(\beta)}(y) \rangle^{DR}_{\rm{amp}}\Big{|}_{m=0}&=&  \frac{1}{\alpha} \delta^{(\alpha)\,(\beta)} q_{\mu} q_{\nu} + \delta^{(\alpha)\,(\beta)}\left(q^2 \delta_{\mu \nu} - q_{\mu} q_{\nu}\right)\times\\\nonumber
&&\Big[ 1  -\frac{g^2\, N_f}{16\,\pi^2} \left( \frac{1}{\epsilon} + 2 +\log\left(\frac{\bar\mu^2}{q^2} \right)\right)\\\nonumber
&& - \frac{g^2\, N_c}{16\,\pi^2}\frac{1}{2} \left(\frac{1+\alpha}{\epsilon} + \frac{7}{2} - \alpha - \frac{\alpha^2}{2} + (1+\alpha)\log\left(\frac{\bar\mu^2}{q^2} \right)\right) \Big]\\[2ex]
\langle  \lambda^{(\alpha)}(x) \bar \lambda^{(\beta)}(y) \rangle^{DR}_{\rm{amp}}\Big{|}_{m=0}&=& \frac{i}{2} \delta^{(\alpha)\,(\beta)}\, \qslash \Big[ 1 - \frac{g^2\,N_f}{16\,\pi^2}\left(1 + \frac{1}{\epsilon} + \log\left(\frac{\bar\mu^2}{q^2} \right)\right)\\\nonumber
&& - \frac{g^2\, N_c}{16\,\pi^2} \left(\alpha + \frac{\alpha}{\epsilon} + 4 \,\alpha \log\left(\frac{\bar\mu^2}{q^2}  \right)\right)\Big]\\[2ex]
\langle  \psi(x) {\cal O}_S(z) \bar \psi(y) \rangle^{DR}_{\rm{amp}}\Big{|}_{m=0}&=&  \openone \left[ 1 + \frac{g^2\,C_F}{16\,\pi^2} \left( \frac{3+\alpha}{\epsilon} + 4 + 2 \alpha  +  (3 + \alpha) \log\left(\frac{\bar\mu^2}{q^2} \right) \right) \right]\\[2ex]
\langle  \psi(x) {\cal O}_P(z) \bar \psi(y) \rangle^{DR}_{\rm{amp}}\Big{|}_{m=0}&=& \gamma_5 \left[ 1 + \frac{g^2\,C_F}{16\,\pi^2} \left( \frac{3+\alpha}{\epsilon} + 12 + 2 \alpha  +  (3 + \alpha) \log\left(\frac{\bar\mu^2}{q^2} \right) \right) \right]\\[2ex]
\langle  \psi(x) {\cal O}_V(z) \bar \psi(y) \rangle^{DR}_{\rm{amp}}\Big{|}_{m=0}&=& \gamma_\mu \left[ 1 + \frac{g^2\,C_F}{16\,\pi^2}  \alpha \left( \frac{1}{\epsilon} + 1  +  \log\left(\frac{\bar\mu^2}{q^2} \right)  \right) \right] - 2 \alpha \frac{q_\mu \qslash}{q^2} \frac{g^2\,C_F}{16\,\pi^2} 
\eea
\be
\langle  \psi(x) {\cal O}_A(z) \bar \psi(y) \rangle^{DR}_{\rm{amp}}\Big{|}_{m=0}= \gamma_5 \gamma_\mu \left[ 1 + \frac{g^2\,C_F}{16\,\pi^2}   \left( \frac{\alpha}{\epsilon} + 4 + \alpha + \alpha \log\left(\frac{\bar\mu^2}{q^2} \right) \right) \right] - 2 \alpha \gamma_5 \frac{q_\mu \qslash}{q^2} \frac{g^2\,C_F}{16\,\pi^2}
\ee
\be
\langle  \psi(x) {\cal O}_T(z) \bar \psi(y) \rangle^{DR}_{\rm{amp}}\Big{|}_{m=0}= \gamma_\mu \gamma_\nu \left[ 1 + \frac{g^2\,C_F}{16\,\pi^2}  \left(\alpha - 1 \right)\left( \frac{1}{\epsilon} +  \log\left(\frac{\bar\mu^2}{q^2} \right) \right) \right].
\label{GF2tensor}
\ee
One can observe that there is no one-loop longitudinal part for the gluon self-energy. Thus the renormalization function for the gauge parameter receives no one-loop contribution. From the above results we can extract the renormalization functions:
\bea
Z_\psi^{DR,\MSbar} &=& 1 + \frac{g^2\,C_F}{16\,\pi^2} \frac{1}{\epsilon}\left( 2 + \alpha \right),\,\,Z_{A_\pm}^{DR,\MSbar} = 1 + \frac{g^2\,C_F}{16\,\pi^2} \frac{1}{\epsilon}\left(1 + \alpha \right)\\[2ex]
Z_{u}^{DR,\MSbar} &=&  1 + \frac{g^2\,}{16\,\pi^2} \frac{1}{\epsilon} \left(\frac{1+\alpha}{2} N_c +  N_f \right),\,\,Z_{\lambda}^{DR,\MSbar} =  1 + \frac{g^2\,}{16\,\pi^2} \frac{1}{\epsilon} \left(4 \,\alpha\, N_c + N_f \right)\\[2ex]
Z_{S, P}^{DR,\MSbar} &=& 1 - \frac{g^2\,C_F}{16\,\pi^2} \frac{1}{\epsilon},\,\,
Z_{V, A}^{DR,\MSbar} = 1 + \frac{g^2\,C_F}{16\,\pi^2} \frac{2}{\epsilon},\,\,Z_T^{DR,\MSbar} = 1 + \frac{g^2\,C_F}{16\,\pi^2} \frac{3}{\epsilon}.
\eea
Calculating the same Green's functions as before on the lattice, and combining them with our results from the continuum, we will be able to extract $Z_\psi^L$, $Z_u^L$, $Z_\lambda^L$, $Z_{A_\pm}^L$ and $Z_\Gamma^L$ in the $\MSbar$ scheme and on the lattice. On the lattice we have to calculate all the diagrams which were presented here as well as further tadpole diagrams containing closed gluon, fermion, ghost loops and for gluon propagator we have also take into account the contribution which comes from the measure part of the action. For the algebraic operations involved in evaluating Feynman diagrams, we make use of our symbolic package in Mathematica.
\section{Future Plans -- Conclusion}
\label{future}
This will be the first calculation of the renormalization functions for SQCD on the lattice, providing a thorough set of results for all counterterms, mixing coefficients and parameter fine-tuning. The determination of the Renormalization Functions of all fields and parameters which appear in ${\cal S}^E_{\rm SQCD}$ and of a complete set of quark bilinear operators on the lattice are necessary ingredients in the prediction of physical probability amplitudes from lattice matrix elements. As a first example of our lattice results, we find for the inverse quark propagator:
\be
\langle \psi^B(x) \bar \psi^B(y) \rangle^{L}_{\rm{amp}}\Big{|}_{m=0}=  i \qslash + \frac{g^2\,C_F}{16\,\pi^2}
\left[ i \qslash \left(12.8025 - 4.79201 \alpha + (2+\alpha)\log\left(a^2\,q^2\right)\right) - \frac{1}{a} 51.4347 \,r  \right] 
\label{GF2quarklatt}
\ee
where $a$ is the lattice spacing.
From Eqs.~(\ref{GF2quarklatt}) and~(\ref{GF2quark}) we extract the renormalization for the quark field on the lattice: $Z_\psi^{L,\MSbar} = 1 + g^2\,C_F/(16\,\pi^2) \left( -16.8025 + 3.79201 \alpha - (2+\alpha)\log\left(a^2\, \bar\mu^2\right) \right)$. In additional the quark critical mass can read from Eq.~(\ref{GF2quarklatt}): $m^{quark}_{crit.} =  -g^2\,C_F/(16\,\pi^2)\,  51.4347 \,(1/a) \,r$.
A further extension of the present work would be to investigate of relationships between different Green's functions involved in SUSY Ward identities and to study the mixing among operators.   

\end{document}